\documentclass [11pt,a4paper] {article}

\usepackage[cp1252]{inputenc}
\usepackage[english]{babel}

\usepackage{amssymb}
\usepackage{amsmath}
\usepackage{amsfonts,amssymb}

\usepackage[dvips]{graphicx}

\usepackage{color}

\begin{document}

\title{Concerning Infeasibility of the Wave Functions of the Universe}

\author{Arkady Bolotin\footnote{$Email: arkadyv@bgu.ac.il$} \\ \textit{Ben-Gurion University of the Negev, Beersheba (Israel)}}

\maketitle

\begin{abstract}\noindent Difficulties with finding the general exact solutions to the Wheeler-DeWitt equation, i.e. the wave functions of the Universe, are known and well documented. However, the present paper draws attention to a completely different matter, which is rarely if ever discussed in relation to this equation, namely, the time complexity of the Wheeler-DeWitt equation, that is, the time required to exactly solve the equation for a given universe. As it is shown in the paper, whatever generic exact algorithm is used to solve the equation, most likely such an algorithm cannot be faster than brute force, which makes the wave functions of the Universe infeasible.\\

\end{abstract}

\section{Introduction}

\noindent Even with the complete lack of observational and experimental evidence that clearly contradicts to either Einstein's theory of general relativity (GR) or quantum mechanics (QM) and, as a result, with no compelling reason to adopt another theory, the aim of describing the quantum behavior of the gravitational field -- called a quantum theory of gravity -- is regarded as one of the biggest tasks of the modern physics. This is probably so because it is widely believed that interplay between GR and QM cannot be avoided if we want to understand how the Universe works \cite{Ashtekar,Kiefer05}.\\

\noindent In the literature, one can find a number of different types of logic rationalizing the necessity of reconciling the laws of QM with GR (see, for example, papers \cite{Albers,Kiefer14} that catalog most of them). Yet, the simplest (and perhaps the most convincing) reason behind the need for quantization of the gravitational field is the quantum superposition principle resulting from the property of linearity of the solutions to the Schrödinger equation -- the central equation of QM. Certainly, if Schrödinger's equation -- established and empirically confirmed for systems made up of a small number of microscopic constituent particles $N$ -- continues to be valid even for systems containing a number of those particles as large as (or even larger than) Avogadro's number ${N}\!_{A} \approx 10^{24}$, then the quantum superposition principle should be applicable to general states of macroscopic systems and, consequently, to the gravitational field produced by macroscopic systems.\\

\noindent Theoretically however, it is quite conceivable that due to certain additional nonlinear stochastic terms present in the Schrödinger equation, the linear nature of QM breaks down in situations where microscopic systems grow into macroscopic ones (i.e., where the number $N$ grows large) and thus the gravitational field that these systems produce becomes distinguishable; see, for example, papers \cite{Bassi,Singh,Penrose,Diosi} that discuss in detail this approach called objective collapse theories or the Dynamical Reduction Program (DRP). However, no empirical hint for such a drastic modification of Schrödinger's equation exists so far. Moreover, a modification of Schrödinger's equation such as this involves phenomenological parameters, which, if the whole DRP is taken seriously, acquire the status of new constants of nature. Therefore, from a pragmatic point of view, it is hard to accept that in order to avoid one theoretical construct, a quantum theory of gravity, we have to agree to another, the DPR.\\

\noindent Withal, from a point of view of computational complexity theory, the stretch of the applicability of the quantum superposition principle far beyond its established microscopic scale is equivalent to the assumption of Schrödinger's equation \textit{scalability} on $N$ -- that is, the assertion that Schrödinger's equation is \textit{solvable in a suitably efficient and practical way} even when it is applied to an arbitrary physical system with large $N$.\\

\noindent Truly, assume the opposite: Schrödinger's equation is not scalable on $N$ and hence there is no practical possibility to solve this equation for an arbitrary system with large $N$ in any reasonable amount of time. In such circumstances, Schrödinger's equation would not be able to actually describe how states of macroscopic systems (possessing $N \gtrsim 10^{24}$ microscopic constituent particles) would change with time. So, in that case, QM as a physical theory would not have the ability to predict for macroscopic systems and, for this reason, the quantum superposition principle could not be practically applicable to the description of states of macroscopic systems.\\

\noindent Thus, if held true, scalability on $N$ of the Schrödinger equation would necessitate the relevance of QM to the description of any macroscopic system including the entire Universe and therefore would lead with necessity to the quantization of the gravitational field. This would imply that it would be possible (albeit in principle) to find the wave functions describing the Universe -- i.e., the general exact solutions to the quantum gravitational analogue of the Schrödinger equation (such as the Wheeler-DeWitt equation) -- in a reasonable amount time. The aim of the present paper is to evaluate such a claim.\\

\section{Quantum model of the Universe}

\noindent For that purpose, let us examine a customary quantum model of the Universe. Following the accepted quantum cosmological doctrine, the quantum state of the Universe is described by a wave function $\Psi[{h}_{ab},\phi]$ (called “the wave function of the Universe”), which is a functional on the three-metric ${h}_{ab}$ (i.e., metric on the hypersurface) and on the values of nongravitational fields symbolically denoted by $\phi$ \cite{DeWitt,Hartle}. The wave function of the Universe $\Psi[{h}_{ab},\phi]$ obeys the Wheeler-DeWitt (WDW) second-order functional differential equation\\

\begin{equation} \label{1} 
   \mathcal{H}\Psi[{h}_{ab},\phi] = 0
\;\;\;\;  ,
\end{equation}
\smallskip

\noindent where  $\mathcal{H}$ stands for the Hamiltonian operator of the Universe (a first class constraint on the Universe physical states) that involves functional derivatives with respect to the metric components ${h}_{ab}$ and the Hamiltonian density $\mathcal{H}_m$ for nongravitational fields $\phi$. The WDW equation basically says that the operator $\mathcal{H}$ acts on the wave functional $\Psi[{h}_{ab},\phi]$, which provides all of the information about the geometry and matter content of the Universe.\\

\noindent For the most part, difficulties with finding the WDW general exact solutions (essential because implications for the meaning of the wave functions of the Universe must be derived from the exact solutions to the WDW equation) are known and well documented (see, for example, the paper \cite{Carlip}, which in detail reviews such difficulties). Here, however, we want to draw attention to a completely different matter, which is rarely if ever discussed in relation to the WDW equation -- the time complexity of this equation, i.e., the time (or the number of elementary operations) required to exactly solve this equation for a given universe.\\

\section{Time complexity of the Wheeler-DeWitt equation}

\noindent Obviously, the amount of time needed to exactly solve the WDW equation is subject to an algorithm used for solving the equation. So therefore, let us figure out how fast such an algorithm might be in principle.\\

\noindent Assume that $A(\Psi[{h}_{ab},\phi])$ is the generic exact algorithm capable of finding the set of the general exact solutions $\Psi[{h}_{ab},\phi]$ to the WDW equation with an arbitrary Hamiltonian constraint $\mathcal{H}$ (i.e., for an arbitrary physical universe with any geometry and any matter content). If this algorithm $A(\Psi[{h}_{ab},\phi])$ were to exist, it would be also capable of exactly solving the Schrödinger equation for an arbitrary matter source. Let us show this.\\

\noindent Consider the ansatz\\

\begin{equation} \label{2} 
   \Psi[{h}_{ab},\phi]
   \equiv
   \exp\left(
        \frac{i}{\hbar}S[{h}_{ab},\phi]
       \right)
   =
   \exp\left(
        \frac{i}{\hbar}\left(
                                     MS_0+S_1+M^{-1}S_2+\ldots
                           \right)
       \right)
\;\;\;\;  ,
\end{equation}
\smallskip

\noindent where $M=(32{\pi}G)^{-1}c^2$ is the parameter, with respect to which the semiclassical expansion is performed in (\ref{2}). Now suppose that the Hamiltonian constraint $\mathcal{H}$ is such that the wave function of the Universe $\Psi[{h}_{ab},\phi]$ is of the special form\\

\begin{equation} \label{3} 
   \Psi[{h}_{ab},\phi]
   =
   \frac{1}{D[{h}_{ab}]}
      \exp\left(
           \frac{i}{\hbar}MS_0[{h}_{ab}]
          \right)
   \psi[{h}_{ab},\phi]
\;\;\;\;  ,
\end{equation}
\smallskip

\noindent where $\psi[{h}_{ab},\phi]$ is the wave functional at the order $M^0$ in the expansion (\ref{2})

\begin{equation} \label{4} 
   \psi[{h}_{ab},\phi]
   =
   D[{h}_{ab}]
      \exp\left(
           \frac{i}{\hbar}S_1[\phi]
          \right)
\;\;\;\;  .
\end{equation}
\smallskip

\noindent Inserting (\ref{3}) into the WDW equation we find that $\psi[{h}_{ab},\phi]$ obeys the functional Schrödinger equation in its local form for quantum fields propagating on the classical spacetimes described by $S_0[{h}_{ab}]$:\\

\begin{equation} \label{5} 
   i\hbar
   \frac{\delta\psi[{h}_{ab},\phi]}{\delta\tau}
   =
   \mathcal{H}_m\psi[{h}_{ab},\phi]
\;\;\;\;  ,
\end{equation}
\smallskip

\noindent where $\tau$ is the time functional (defined on the configuration space) generating the time parameter $t$ in each classical spacetime (see \cite{Kiefer94} for a review of the semiclassical approximation to quantum gravity in the canonical framework).\\

\noindent Unlike the Schrödinger equation for a $N$-particle system, where a wave function $\psi(\phi)$ evolves against a classical background potential $U(\phi)$, the functional equation (\ref{5}) is a ``second-quantized'' equation, therefore its eigenstate $\psi[{h}_{ab},\phi]$ is a wave functional on the configuration space, whose points are field configurations. Nevertheless, in the non-relativistic limit and for the negligible gravitational field ($c=\infty$ and $G=0$), the field Hamiltonian ${H}_m$ will take the form of an expression for the expectation value of the $N$-particle system's energy, with $\psi[{h}_{ab},\phi]$ playing the role of the wave function $\psi(\phi)$ of the system. It follows, then, that in the limits $c=\infty$ and $G=0$, by finding the solution $\Psi[{h}_{ab},\phi]$ to the WDW equation, the generic algorithm $A(\Psi[{h}_{ab},\phi])$ yields the solution $\psi(\phi)$ to the Schrödinger equation as well.\\

\noindent When evaluating the difficulty of a particular Hamiltonian, the amount of computation (measured as the time or the number of elementary operations required to verify the energy of the ground state of the given $N$-particle system) defines the complexity of the Hamiltonian. If for the given system the ground state energy can be verified in the time polynomial in $N$, then the system's Hamiltonian is in the NP complexity class (i.e., in the class of computational problems whose solutions can be verified in polynomial time). If a particular Hamiltonian has enough flexibility to encode any other problem in the NP class, then the Hamiltonian is considered NP-complete.\\

\noindent Take, for example, the Ising Hamiltonian function $H\!\left({\sigma}\!_1,\!\dots\! ,{\sigma}\!_j,\!\dots\! ,{\sigma}\!_{N}\!\right)$ describing the energy of configuration of a set of $N$ spins $\sigma\!{_j}\hbar=2\phi\!{_j}\in\left\{\!-\hbar, +\hbar\right\}$ in classical Ising models of a spin glass:\\

\begin{equation} \label{6} 
   H\!\left({\sigma}_1,\dots ,{\sigma}_j,\dots ,{\sigma}_N\!\right) 
   =
  -\sum_{j<k}{A_{jk}\sigma\!{_j}\sigma\!{_k}}
  -B\sum^N_{j}{C_{j}\sigma\!{_j}}
  \;\;\;\;
  \left(A_{jk},B,C_{j}=\rm{const} \right)
\;\;\;\; .
\end{equation}
\smallskip

\noindent Considering that all problems in the NP class can be mapped to the Schrödinger equation with the quantum version of the Ising Hamiltonian $H\!\left({\sigma}\!_1,\!\dots\! ,{\sigma}\!_j,\!\dots\! ,{\sigma}\!_{N}\!\right)$, in which spins $\sigma\!{_j}$ have been merely replaced by quantum operators, Pauli spin-1/2 matrices ${\sigma}_j^z$,\\

\begin{equation} \label{7} 
   H\!({\sigma}_1^z,\dots ,{\sigma}_j^z,\dots ,{\sigma}_N^z\!)
   \psi(\phi)
   =
  0
\;\;\;\; ,
\end{equation}
\smallskip

\noindent the Hamiltonian $H\!({\sigma}_1^z,\dots ,{\sigma}_j^z,\dots ,{\sigma}_N^z\!)$ is NP-complete (see paper \cite{Lucas}, which provides Ising formulations for many NP-complete and NP-hard problems, including all of Karp's 21 NP-complete problems, for details).\\

\noindent In this way, we can conclude that as long as the generic algorithm $A(\Psi[{h}_{ab},\phi])$ can solve the Schrödinger equation recovered, in the weak field approximation, from the WDW equation, this algorithm can resolve any NP-complete problem.\\

\noindent On the other hand, every NP-complete problem can be exactly solved by \textit{brute force}, that is, by the generic and exact algorithm of enumerating and checking all possible candidate solutions to the problem. Obviously, brute force is not efficient: For example, deciding whether there is a spin configuration $\phi=({\phi}_1,\dots ,{\phi}_j,\dots ,{\phi}_N\!)$ of a spin glass with zero energy by exhaustively checking all possible spin configurations may well require $O^*(2^N)$ elementary operations (where the $O^*$ notation is used that suppresses all factors polynomial in $N$).\\

\noindent Consequently, the question becomes, is it possible that the generic exact algorithm $A(\Psi[{h}_{ab},\phi])$ can be significantly faster than brute force? The short answer is no, according to what now seems true, it is not possible.\\

\noindent Indeed, assume that the algorithm $A(\Psi[{h}_{ab},\phi])$ is a sub-exponential time algorithm (i.e., faster than brute force). It follows then $A(\Psi[{h}_{ab},\phi])$ can solve any NP-complete problem in sub-exponential time. This means that many computational problems known to be solved in exponential time $O^*(2^N)$ can be improved to $O^*(c^N)$ with some $c<2$. However, such an improvement would be highly surprising since it would refute the Strong Exponential Time Hypothesis (SETH) -- the conjecture based on evidence that the low bond $O^*(2^N)$ matches the running time of the best possible algorithms for many computational problems (see \cite{Woeginger,Lokshtanov} for the survey of results obtained in the field of exact exponential time algorithms under the assumption of the SETH).\\

\noindent Therefore, almost certainly no generic algorithm for finding the general exact solutions to the WDW equation can be significantly faster than brute force. So, it is extremely probable that the wave functions of the Universe are infeasible, i.e., they cannot be calculated in any reasonable time.\\

\section{Concluding remarks}

\noindent Let us dwell for a moment on this conclusion trying to think about what it means.\\

\noindent Does it mean that even if we manage to overcome all the difficulties arise in merging QM with GR and finally build a definitive self-consistent theory of quantum gravity, we will know nothing more about the Universe, gravitating bodies, the motion of planets and stars than before having this theory since it will not be able to make practically meaningful predictions?\\

\noindent Paradoxically, the answer should be yes (the only way to dodge such an answer would be equality of the NP class to the P class of computational problems solvable in polynomial time; yet, this equality is prevalently believed to be not true).\\

\noindent One can make an objection here that even though we cannot exactly solve differential equations such as the Schrödinger equation except for some simple cases, it still does not mean that those equations should be stripped of any physical significance. There are many well-defined approximation schemes (such as the Born-Oppenheimer approximation) from which one can derive concrete, experimentally testable, results. So, why this should be different for the WDW equation?\\

\noindent Certainly, it is true that many features of quantum cosmology can be discussed using the approximate solution of the WDW equation (i.e., in the limit when this solution assumes a semiclassical or WKB form). It has been even suggested that the wave function of the universe should be interpreted only in the WKB limit, because only then a time parameter and an approximate (functional) Schrödinger equation is available (see for example paper \cite{Vilenkin}). However, as it was observed in \cite{Kiefer14}, implications for the meaning of the quantum cosmological wave functions should be derived as much as possible from exact solutions to the WDW equation, otherwise such implications can lead to a conceptual confusion. This is because the WKB approximation breaks down in many interesting situations, even for a universe of macroscopic size. One example is a closed Friedmann universe with a massive scalar field \cite{Kiefer88}, and one more example is the case of classically chaotic cosmologies (see \cite{Calzetta} and \cite{Cornish} for detail).\\

\noindent Another objection may raise that if we followed the presented in the paper 'computational' argumentation, we would not be able to understand the success of quantum theory. There, we use functional differential equations or, alternatively, path integrals. These equations are in general not exactly soluble, but we can derive from them physically meaningful results. Again, why the situation with the WDW equation should be different?\\

\noindent Since quantum mechanics was introduced, its most insight has been gained from understanding the exact (or quasi-exact) solutions to the Schrödinger equation. Such solutions are very important because they convey maximum information of a system. Besides, these solutions are valuable tools in checking and improving models and numerical methods being introduced for solving complicated physical problems. From computational complexity perspective, those solutions were possible to find in reasonable time because of either small inputs of the problems (such as in the cases of physical systems composed of a few constituent particles completely isolated from the environment) or the presence of system-specific heuristics that could be used to drastically reduce the complexity of the system's Schrödinger Hamiltonian (i.e., the cardinality of the set of all possible candidates for the witness of the given system).\\

\noindent For example, suppose a macroscopic system can be formally divided into a 'collective' system represented by a small set of the system's macroscopic observables (along with their conjugate partners) correspond to properties of the macroscopic system as a whole and the environment, which is the set of the system's observables other than the ‘collective’ ones. Since the microscopic degrees of freedom of an ordinary macroscopic system are uncontrolled for the most part, one cannot hope to keep track of all the degrees of freedom of the environment. This may be used as a heuristic allowing an enormous set of all possible candidate solutions for the macroscopic system to be reduced to just a small set comprising only candidate solutions for the ‘collective’ system. Upon applying this heuristic by way of “tracing out” the uncontrolled degrees of freedom of the environment and assuming that the environmental quantum states are orthogonal, one would get an inexact yet practicable solution to Schrödinger's equation approximately identical to the corresponding mixed-state density matrix of the ‘collective’ system describing the possible outcomes of the macroscopic observables of the macroscopic system and their probability distribution. As one can readily see, the above-described heuristic represents the decoherence process that transforms a pure quantum state into a mixed state.\\

\noindent However, the question is what can be considered as 'background (environmental) variables' for the WDW equation. Except the cases where a modeled system can be treated as an open system or where the irrelevant (or negligible) degrees of freedom may include tiny gravitational waves and density fluctuations, there is no clear and general answer to this question.\\

\noindent The conclusion about infeasibility of the wave functions of the Universe can also challenge the meaning of a solution to a differential equation of a physical theory. Usually getting a solution means the ability to predict, that is, to calculate the value of the physical system's quantity. On the other hand, if solutions to the WDW equation are infeasible, they would be practically useless (they cannot be used to calculate values of the Universe) and accordingly they can add nothing to our understanding about how the Universe works.\\

\noindent Therefore, does this imply that unfeasible, i.e., practically useless, solutions to differential equations of physical theories should be defined as unphysical and thus excluded from consideration?\\

\noindent Clearly, this is a very difficult question disturbing the fundamentals of physical science. However, without a comprehensive answer to that question the further progress in quantum gravity, in all likelihood, seems impossible.\\

\section*{{Acknowledgments}}

\small{\noindent The author wishes to thank the anonymous reviewer who has greatly assisted him in improving this paper.}\\

\end{document}